\documentclass[twocolumn,preprintnumbers,amsmath,amssymb]{revtex4}

\usepackage{graphicx}% Include figure files
\usepackage{dcolumn}% Align table columns on decimal point
\usepackage{bm}% bold math
\begin{document}

\title{Fe adatoms along Bi lines on H/Si(001): Patterning atomic magnetic chains}

\author{W. Orellana$^1$ and R. H. Miwa$^2$}

\affiliation{$^1$Departamento de F\'{\i}sica, Facultad de Ciencias, 
Universidad de Chile, Casilla 653, Santiago, Chile\\
$^2$Instituto de F\'{\i}sica, Universidade Federal de Uberl\^andia, 
C.P. 593, CEP 38400-902, Uberl\^andia, MG, Brazil}

%date\today

\begin{abstract}
The stability, electronic and magnetic properties of Fe atoms adsorbed on 
the self-assembled Bi-lines nanostructure on the H/Si(001) surface are 
addressed by spin-density functional calculations. Our results show that 
Fe adatoms are much more stable on sites closer to the Bi nanolines, suggesting
that they form one-dimensional atomic arrays. The most stable structure occurs 
on a missing dimer line beside the Bi dimers, which corresponds to an array  
with distances between Fe adatoms of about 8~\AA. In this array the irons 
are coupled antiferromagnetically with spin magnetic moment of about 
1.5~$\mu_{\rm B}$ per Fe atom, whereas the coupling exchange interactions is 
found to be of 14.4~meV. We also estimate a large magnetic anisotropy energy 
of about 3~meV/atom originated on the structural anisotropy of the Fe-adatom
site. In addition, we find that the electronic band structure of the Fe array 
at the most stable structure shows a magnetic half-metal behavior. 
\end{abstract}

%pacs{75.75.+a,73.22.-f,68.43.Fg}

\maketitle 
Self-organized nanostructures on surfaces have attracted
much attention during the last few years owing their promising
applications in the patterning of low-dimensional magnetic systems
\cite{fruchart}. Suitable epitaxial techniques make possible
to build one-dimensional (1D) arrays of $3d$ magnetic atoms by step
decoration of metallic substrates, which is currently a very active
topic of research in magnetism from both experimental
\cite{shen,pratzer,gambardella1} and theoretical
\cite{dorantes,spisak} viewpoint. New fundamental physical phenomena
like magnetic ordering and giant magnetic anisotropy have been
observed in self-assembly 1D arrays of magnetic atoms
\cite{gambardella2}.  On the other hand, little is known about the
spin interactions of magnetic atoms in contact with semiconducting
surfaces. Recently, thin films of magnetic transition metal on Si(001)
have been studied from first principles by Wu {\it et al.}  \cite{wu}.
They found that MnSi films on Si(001) are ferromagnetic with sizable
magnetic moment, whereas FeSi and NiSi are nonmagnetic.  Following up
on earlier studies of magnetism in monoatomic arrays grown by self
assembly on metallic substrates, we explore similar magnetic arrays on
semiconducting surfaces taking advantage on the remarkable structural
quality of the self-assembled Bi-dimer lines on the Si(001) surface.
These nanolines, which are obtained by Bi deposition onto Si(001)
above the desorption temperature of 500~$^{\circ}$C, consists of two 
parallel rows of symmetric Bi dimers which are about 0.6~nm apart and
can be over 500~nm in length. Additionally, their structures are free 
of defects like kinks or breaks and have a remarkable straightness 
\cite{naitoh,miki,owen0}. However, possible template applications require 
the hydrogen exposure of the Si(001) substrate, since it is more 
reactive than the Bi line. After hydrogenation of Bi/Si(001), the H 
atoms only terminate the Si ones leaving the Bi lines clean and 
preserving their 1D structures \cite{naitoh,owen1}. Recently, the 
arrengement of nonmagnetic atoms on the Bi-nanolines structure as well 
as their use as a template have been experimentally investigated 
\cite{owen2,bowler,owen3}.

In this work we explore the possibility to construct magnetic monoatomic
chains of Fe adatoms by decoration of self-assembled Bi lines
on H/Si(001) surfaces.  We find that the Fe adatoms have the
most-stable structure beside the Bi-dimer, forming nearly 1D
atomic arrays following the Bi-dimer lines with a Fe-Fe distance of
about 8~\AA. The Fe adatom array in the most stable configuration is
antiferromagnetic, having a weak exchange coupling of 14.4~meV,
suggesting the formation of extremely narrow magnetic domain walls.
We also estimate a lower limit for the magnetic anisotropy energy
(MAE) which is the energy involved in rotating the magnetization 
from a direction of low energy toward one of high energy. We find 
a very large MAE of about 3~meV/atom, which suggest a relatively 
high energy barrier to change the magnetization from preferential
directions.
Concerning the electronic properties, the 1D Fe array shows a
magnetic half-metal behavior, i.e., the majority-spin electrons are
semiconducting whereas the minority-spin electrons are metallic. The
above results turn the Fe arrays on Bi lines an interesting system
both for basic research and for possible technological applications,
for instance in spintronics and nanoscale data-storage devices. We
hope that experimental studies will be motivated by our predictions.

The calculations were performed in the framework of the density functional 
theory within the local-spin-density approximation (LSDA)~\cite{perdew}, 
considering non-collinear magnetic ordering as implemented in the SIESTA 
code~\cite{siesta}. The basis set consists of numerical pseudoatomic orbitals,
namely, double-$\zeta$ plus polarization functions. Standard norm-conserving 
pseudopotentials in their separable form were used to describe the 
electron-ion interaction, including nonlinear core correction for Fe and 
Bi atoms \cite{louie}. Currently, there are two energetically favorable structural 
models to explain the observed self-assembled Bi-dimer lines on clean and 
hydrogenated Si(001) substrates. These are the Miki\cite{miki} and the 
Haiku\cite{owen0} structures. We adopt the latter one because it exhibits 
the highest stability on H/Si(001) as recently shown by {\it ab initio} 
calculations \cite{miwa}. To simulate the Bi lines in the Haiku model structure 
we used the repeated slab method within a 2$\times$6 surface unit cell, 
containing ten monolayers of Si atoms plus a vacuum region of about 11~\AA. 
The bottommost Si dangling bonds were saturated by hydrogens. Four special 
{$k_\parallel$} points were used to sampling the surface Brillouin zone. 
The topmost eight monolayers were fully relaxed until the force components 
become smaller than 0.05~eV/\AA.

\begin{figure}
\includegraphics[width= 7.0cm]{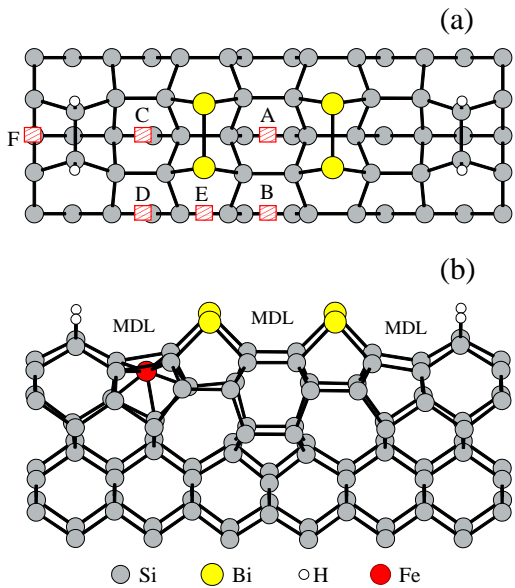}
\caption{(Color on line) (a) Top view equilibrium geometry for the 
clean Bi line on H/Si(001). Dashed squares indicate the Fe adsorption 
sites under consideration. (b) Most stable equilibrium geometry 
resulting from the Fe adsorption on the site $C$.}
\label{f1}
\end{figure}

We investigate six adsorption sites for the Fe atom on the Bi-dimer structure 
formed on the hydrogenated Si(001) substrate (hereafter the clean Bi line), as 
shown in Fig.~\ref{f1}(a). In the clean Bi line, two parallel Bi dimers align 
along the [110] direction which are separated by a missing dimer line (MDL). 
In Addition, two adjacent MDLs form beside the Bi dimers. The Bi-dimer bond 
lengths and the lateral distance between them are found to be 3.15 and 6.36~\AA, 
respectively. Our results for the equilibrium geometry of the clean Bi line is 
in good agreement with previous plane-waves {\it ab initio} results~\cite{miwa}.
%%%%%%%%%%%%%%%%%%%%%%%%%%%%%%%%%
%%%%%%%%%%%%%%%%%%%%%%%%%%%%%%%%%

Figure~\ref{f1}(b) shows the equilibrium geometry for the most stable
structure of the Fe adatom at the MDL beside the Bi dimers ($C$
site). We find that the Fe atom tends to increase the number of Fe-Si
bonds occupying an interstitial subsurface position, about 0.5~\AA\
below the topmost Si atoms, becoming sevenfold coordinated with bond
lengths ranging from 2.2 to 2.5~\AA. Similar coordination numbers are
found for the Fe adsorption on the other sites. The binding energy of
the Fe adatom on the $C$ site is found to be about 6~eV. These results
somewhat mimic the B20 phase of FeSi iron silicide. The Fe atoms in
FeSi (B20) have seven nearest-neighbor Si atoms, and binding energy of
6.58~eV, obtained by LSDA calculations \cite{moroni}. Table~\ref{t1}
shows relative total energies for the Fe adatom at each sites. We note
that the Fe adatom on sites $A$, $B$, and $D$ are close in energy that
the most stable one. This suggests that the MDLs are the energetically
favorable positions for the Fe adsorption.  On the other hand, Fe
adatom between Bi dimers of the same line and on the hydrogenated
Si(001) far from the Bi dimers, sites $E$ and $F$ respectively, are
0.82 and 1.07~eV higher in energy.
%%%%%%%%%%%%%%%%%%%%%%%%%%%%%%%%%
 \begin{table}[b]
 \caption{\label{t1} Total energy ($\Delta E$) with respect to the most
  stable structure ($C$ site), binding energy ($E_b$) of the Fe adatom
  and spin magnetic moment ($m$) of ferromagnetically coupled Fe arrays.
  CN is the Fe-adatom coordination number.}
 \begin{ruledtabular}
 \begin{tabular}{ccccc}    
  Sites & $\Delta E$ (eV) & $E_b$ (eV/Fe) & $m$ ($\mu_B$/Fe) & CN\\
 \hline
   $A$  &     0.134       &    5.834      &    1.83   & 6 \\
   $B$  &     0.350       &    5.618      &    1.63   & 6 \\
   $C$  &     0.000       &    5.968      &    1.53   & 7 \\
   $D$  &     0.120       &    5.848      &    1.27   & 7 \\
   $E$  &     0.816       &    5.152      &    1.99   & 8 \\
   $F$  &     1.072       &    4.896      &    1.95   & 7 \\
 \end{tabular}
 \end{ruledtabular}
 \end{table}
%%%%%%%%%%%%%%%%%%%%%%%%%%%%%%%%%  
%%%%%%%%%%%%%%%%%%%%%%%%%%%%%%%%%
\begin{figure}
\includegraphics[width= 7.5cm]{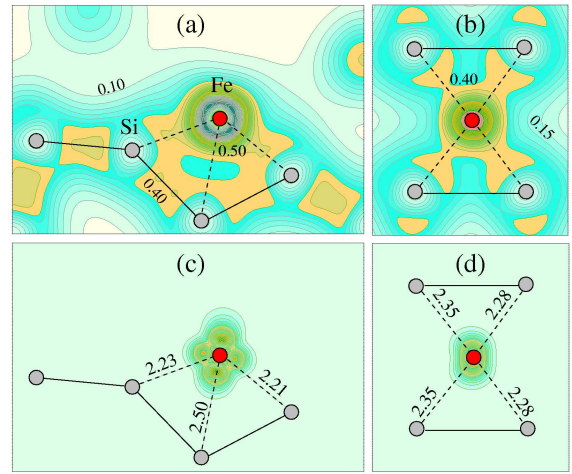}
\caption{(Color on line) (a) and (b) Total charge density on a plane 
perpendicular to the Bi dimers passing by the Fe atom, and through a 
plane passing by the topmost four Si atoms nearest neighbors to the 
Fe atom, respectively. (c) and (d) show the magnetization through the 
same planes than (a) and (b). Numbers in the upper panel correspond 
to the contour charge density in $e$/\AA$^3$, and those of the lower 
panel correspond to bond lengths in \AA.}
\label{f2}
\end{figure}
%%%%%%%%%%%%%%%%%%%%%%%%%%%%%%%%%
At the $E$ site, the Fe adatom breaks the
Bi dimers, forming two Fe--Bi bonds with lengths of 2.58~\AA\ and four 
Fe--Si bonds with lengths of 2.2 and 2.7~\AA. This is the only case where 
the Bi dimers break. We also find that all Fe adatom structures are 
magnetic with magnetic moment ranging from 1.5 to 2.0~$\mu_{B}$ per Fe
atom. The above results suggest that the Bi lines would be effective to 
adsorb Fe atoms, supporting the proposed formation of 1D magnetic
arrays. It is important to note that the energetic preference for the 
adsorption sites $A$ to $D$ indicates that the Bi-line structure would not 
be destroyed by the Fe adsorption. 

Figures~\ref{f2}(a) and \ref{f2}(b) show total charge densities for the 
Fe adatom on the site $C$ on planes perpendicular to the Bi dimers passing 
by the Fe atom, and parallel to the surface passing by the topmost four Si 
atoms nearest neighbors to Fe, respectively. Here we see that the iron 
binds with their seven silicon nearest neighbors, forming metallic bonds.
Typically, pairs of Fe-Si bonds have same lengths whereas an extra one is 
longer, showing a highly anisotropic environment. Similar results are found 
for the other Fe adsorption sites. 
High-coordinated Fe adatoms at subsurface has been also verified in several 
studies addressing the formation of iron silicide thin films on 
Si(001)~\cite{profeta,wu} and Si(111)~\cite{walter,junqueira}. 

\begin{figure}[t]
\includegraphics[width= 7.5cm]{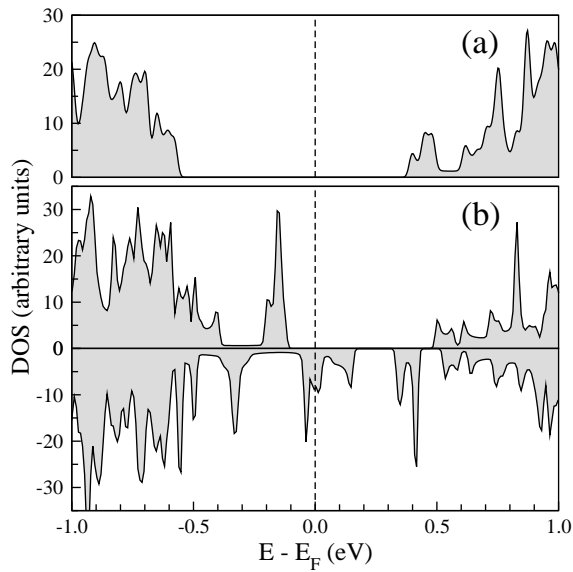}
\caption{(a) Density of states (DOS) for the clean Bi line on H/Si(001). 
(b) Spin-resolved DOS for the Fe adatom on the Bi lines at the most stable 
structure.}
\label{f3}
\end{figure}

\begin{figure}[t]
\includegraphics[width= 8.0cm]{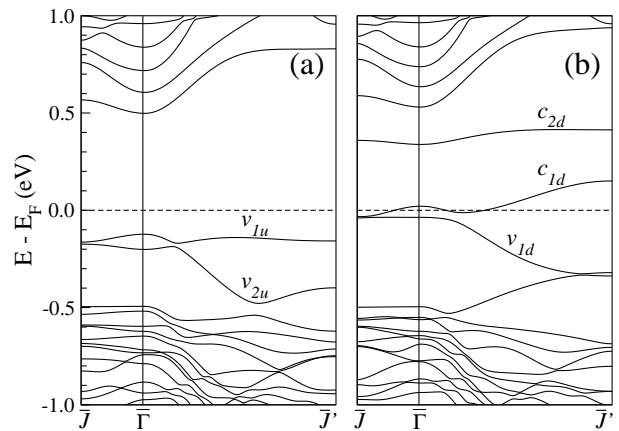}
\caption{Electronic band structure for the Fe adatom on the Bi lines at most 
stable structure. (a) and (b) indicate the majority and minority spin bands, 
respectively.}
\label{f4}
\end{figure}

In Figure~\ref{f3}(a) we present the calculated density of states (DOS) for 
the clean Bi line on H/Si(001). We find that this system is semiconducting 
exhibiting an energy gap of about 1~eV. The occupied Bi states lie resonant 
close to the top of the valence band, whereas the electronic states attributed 
to the Si-dimer dangling bonds are suppressed by the hydrogen saturation,
opening the energy gap.
%%%%%%%%%%%%%%%%%%%%%%%%%%%%%%%%%
%%%%%%%%%%%%%%%%%%%%%%%%%%%%%%%%%
Figure~\ref{f3}(b) shows the spin-resolved DOS for the Fe adatom on
the Bi line in the most stable structure, which reveals different
electronic properties for each spin channels. While the majority spin
exhibits a semiconducting character, the minority spin appears to be
metallic. In order to elucidate these findings, we plot the electronic
band structure for both majority and minority spin channels, as shown
in Figs.~\ref{f4}(a) and \ref{f4}(b), respectively. In these figures,
the $\bar\Gamma\bar J'$ ($\bar\Gamma\bar J$) direction corresponds to
wave vectors parallel (perpendicular) to the Bi lines. It is worthy to
note that the gap states dispersion along $\bar\Gamma\bar J$ are
rather artificial, originating in the supercell construction.  The
distance between neighboring Bi lines in our calculation is about
17~\AA, however the experimentally observed are much more separated.
We find that the electronic states $v_{1u}$, $v_{2u}$, $v_{1d}$, and
$c_{2d}$ come from the $3d$ orbitals of Fe adatom, with negligible
contribution from Si and Bi atoms.
The spin-resolved density of states for the structural model C,
Fig.~\ref{f3}(b), indicates that the net magnetization of Fe arrays
comes mainly from the electronic states induced by Fe adatoms at
$E_{F} - 0.15$~eV ($v_{1u}$), with a small contribution from the
electronic states lying at $E_{F} - 0.4$~eV ($v_{2u}$).  On the other
hand, the minority spin density exhibits two peaks, one at $E_{F}
-0.32$~eV, and another one crossing Fermi level.  The latter
contribution comes from $v_{1d} + c_{1d}$, see Fig.~\ref{f4}(b).  We
observe that the $c_{1d}$ state gives a semimetallic character in the
minority spin channel along the Bi line, however, the majority spin
channel behaves as a semiconductor with a bandgap of about
0.6~eV. This behavior, known as magnetic half metals, was initially
predicted by band structure calculations in Heusler alloys
\cite{groot}. Similar magnetic half-metal behavior is found for the 
Fe adatom at the A site. However, for the Fe adatom at the $B$ and 
$D$ sites, both spin channels have a metallic character. The above
results also suggest that Fe adatoms on Bi lines would be a reliable
system for spin injection from a ferromagnetic material into a
semiconductor. According to our results, if an spin-polarized current
is successfully injected into the most stable Fe array, only one spin 
channel would be able to initiate de conduction. 
%%%%%%%%%%%%%%%%%%%%%%%%%%%%%%%%%
%%%%%%%%%%%%%%%%%%%%%%%%%%%%%%%%%

\begin{figure}
\includegraphics[width= 8.0cm]{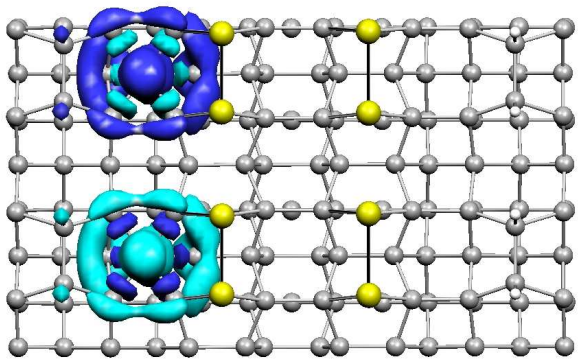}
\caption{(Color on line) Isosurfaces of the antiferromagnetic magnetization 
induced by the Fe adatoms along the Bi lines in the most stable structure. The 
clear (dark) isosurfaces correspond to the majority spin (minority spin) 
density of $+0.004$ ($-0.004$)~$e$/\AA$^3$.}
\label{f5}
\end{figure}

Figures~\ref{f2}(c) and \ref{f2}(d) show contour plots for the net
magnetization, defined as 
$m({\bf r})=\rho_{\uparrow}({\bf r}) - \rho_{\downarrow}({\bf r})$, 
where $\rho_{\uparrow}$ and $\rho_{\downarrow}$ represents the total 
charge density for the majority and minority spin channels, respectively. 
In order to establish the strength of magnetic coupling ($J$) we calculate 
the difference in energy of the ferromagnetic (FM) and antiferromagnetic 
(AFM) ordering by doubling the supercell size. Our results show that the 
Fe adatoms at the $C$ site are coupled antiferromagnetically with a weak 
exchange interaction of $J=14.4$~meV. 
The presence of nonmagnetic Si substrate intermediating the Fe adatoms 
suggests a superexchange origin for the AFM coupling. Figure~\ref{f5} 
shows isosurfaces of the AFM magnetization calculated within the double 
supercell scheme. We observe that the Fe adatom polarizes with an opposite 
spin density the electronic charge between the Si-Fe bonds. The only 
atoms which are effectively polarized are the hydrogens nearest neighbors. 
Magnetic moments of the ferromagnetically coupled Fe atoms in the different 
adsorption sites are shown in Table~\ref{t1}. We note that the highest 
magnetizations occur for the configuration $E$. Here, the Fe adatom binds 
with the Bi atoms, breaking the dimer. This suggests that the increased 
magnetization are due to the polarization of the Bi atoms which contribute 
with additional electrons when forming the Fe-Bi bonds. For the Fe adsorption 
on the H-saturated Si dimers, site $F$, we find a magnetic moment of 
1.95~$\mu_B$ per Fe atom, in close agreement with a previous calculation 
\cite{kishi}.
%%%%%%%%%%%%%%%%%%%%%%%%%%%%%%%%%
%%%%%%%%%%%%%%%%%%%%%%%%%%%%%%%%%

We find that the structural anisotropy of the Fe adatom site induces
a magnetic anisotropy which would be originated in local magnetic dipolar
interactions. We estimate a lower limit for the magnetic anisotropy 
energy by taking the difference in energy for the spin
magnetization pointing in a direction of low energy (easy axis) toward
one of high energy (hard axis). By non-collinear spin calculations we
find that the easy axis occurs at $\theta \approx 0^{\circ}$ and
$\theta \approx 117^{\circ}$, whereas the hard axis occurs at
$\theta \approx 50^{\circ}$, where $\theta$ is the azimuth angle. In
both cases, the magnetization vector is parallel to the Bi dimers. We
find a very large MAE of about 3~meV/atom without considering
spin-orbit contributions, which are not included in the present
calculations. Although the orbital moment would be important for the
magnetic anisotropy of adatoms on metallic surfaces, for instance Co
adatom on Pt(111) \cite{gambardella2}, it would be small or even
quenched for Fe adatom on the Bi-lines structure due to its enhanced 
coordination, which resembles a bulk-like environment. 

In summary, we have examined the stability, electronic and magnetic
properties of Fe atoms adsorbed on the neighborhood of the Bi-dimer 
nanolines on H/Si(001), using spin-density functional calculations. 
Our results show that the Fe atoms tend to occupy highly coordinated 
subsurface position beside the Bi lines, suggesting the formation of
nearly 1D atomic arrays. At the most stable configuration the Fe array 
couples antiferromagnetically with a weak exchange coupling of 14.4~meV. 
We estimate a large magnetic anisotropy energy for a single Fe adatom 
of about 3~meV/atom, suggesting a relatively high energy barrier to 
change the magnetization from the preferential directions. In addition,
the electronic band structure of the most stable Fe array shows a 
magnetic half-metal behavior for the spin channels. Considering that
Bi nanolines on Si(001) is currently grown by self assembling, we have 
shown that its use as a pattern in the design of nearly 1D magnetic 
arrays is energetically favorable. Our results suggest possible 
applications in nanomagnetic devices as well as in spintronics, 
requiring complementary experimental investigations.

%\acknowledgments
This work was supported by the Chilean agency FONDECYT, under Grant Nos. 
1050197 and 7050159. W.O. acknowledges the Millennium Nucleus of Applied 
Quantum Mechanics and Computational Chemistry for financial support, through
Project No. P02-004-F. R.H.M. acknowledges the Brazilian agencies CNPq and 
FAPEMIG.

\end{document}